\documentclass[12pt]{article}
\usepackage{amssymb}
\usepackage{graphicx}
\usepackage[cp1251]{inputenc}
\usepackage{rotating}
\begin{document}
   \bigskip

 \bigskip
 \centerline{\bf\Large Search for a Globular Cluster whose Passage through}
 \centerline{\bf\Large the Galactic Disk could Trigger the Radcliffe Wave}
\bigskip
 \bigskip
 \centerline{\bf
            V. V. Bobylev\footnote [1]{E-mail: bob-v-vzz@rambler.ru},
            A. T. Bajkova
            }
 \bigskip
   \centerline {\small \it Main (Pulkovo) Astronomical Observatory of the Russian Academy of Sciences, St. Petersburg, Russia}
 \bigskip
 \bigskip
Using a catalog of 152 globular clusters (GCs), their orbits were constructed to determine their intersections with the Galaxy's plane of symmetry. Young open star clusters (OSCs) were selected from the selection zone characteristic of the Radcliffe wave. The Hunt and Reffert catalog served as the source of data on these OSCs. A grouping of 17 OSCs with an average age of 32.7 million years was found. It is compact in coordinate, velocity, and age space. This grouping is shown to be a good candidate for the hypothesis that the Radcliffe wave is generated by the passage of an impactor through the Galaxy's plane of symmetry, with the impactor being the GC NGC~4372. The last time it crossed the galactic plane was 55.5 million years ago, and 22.2 million years later, a burst of star formation occurred at this location, forming a whole group of open-clustered stars, and possibly the Radcliffe wave as a whole.

\bigskip\noindent
{\it Keywords:} star formation, globular clusters, NGC~4372, Radcliffe Wave, open star clusters

 \newpage
 \section{INTRODUCTION}
The Radcliffe wave was discovered by Alves et al. (2020) from an analysis of a large sample of molecular clouds. It is a narrow chain of clouds, $\sim$2.7~kpc long in the galactic $XY$ plane, having an inclination to the $Y$ axis of about $30^\circ$. Its main feature is the wave-like nature of the cloud distribution in the vertical direction with a maximum value of the $z\sim160$~pc coordinate in the immediate vicinity of the Sun. A general radial motion of the wave towards the galactic anticenter was detected, as well as motion in the direction of galactic rotation (Konietzka et al. 2024). A wave-like nature of vertical velocities in the Radcliffe wave was established (Bobylev et al. 2022; Konietzka et al. 2024).

Various authors have proposed a number of hypotheses regarding the causes of the Radcliffe wave, none of which is currently generally accepted. Possible hypotheses discussed include, for example, the development of a Kelvin-Helmholtz instability in the galactic disk due to the difference in rotation velocities between the dark matter halo and the disk (Fleck 2020), the influence of the Parker instability of the galactic magnetic field as the cause of the formation of wave-like inhomogeneities in the galactic disk (Bobylev et al. 2025), and the impact of an external impactor, such as a dwarf galaxy satellite of the Milky Way, on the galactic disk (Thulasidharan et al. 2022). Other proposed models include the influence of shock waves from multiple supernova explosions and their stellar winds on the structure of the galactic disk, initiating, for example, the formation of the Local Bubble or the North Polar Spur (Marchal and Martin 2023; Konietzka et al. 2024). It is worth noting that the model involving an external impactor as the trigger for the Radcliffe wave is the most frequently discussed.

A globular cluster (GC) crossing the galactic disk can stimulate star formation. This can result in a) gravitational focusing, where the approach to the GC disk leads to the compression of disk matter toward a specific point, and/or b) strong compression of disk matter in a specific direction. According to calculations by Levy (2000), the passage of a GC through the galactic disk leads to the generation of a shock wave. Based on a model of gravitational focusing of gas, Wallin et al. (1996) concluded that large OB associations can form approximately 30 million years after a globular cluster crosses the disk.

The modeling of the passage of a GC through the galactic disk with the subsequent formation of an open star cluster (OSC) is the subject of the works of Brosche et al. (1991), Rees and Cudworth (2003), Vande Putte and Cropper (2009), de la Fuente Marcos et al. (2014), Bobylev and Bajkova (2018; 2019).

The aim of this paper is to re-evaluate the frequency of globular cluster passages through the galactic plane using the method of numerical integration of orbits. Using current data, we will search for possible cases of the formation of open-clusters in the galactic disk triggered by the passage of a GC. Most importantly, we will search for a globular cluster whose passage through the galactic disk could trigger the emergence of a Radcliffe wave.

 \section{DATA}
We used the Hunt and Reffert (2024) catalog as a source of data on open star clusters. Kinematic and photometric characteristics of stars were taken from the Gaia DR3 catalog (Gaia Collab 2023). The Hunt and Reffert (2024) catalog includes 5647 open clusters and 1309 moving groups. Most of them are located at distances from the Sun less than 4 kpc. The catalog provides the average values of the open cluster trigonometric parallaxes, proper motions, and radial velocities. Age estimates of the clusters, obtained by these authors using the isochrone method, are also given.

\begin{figure}[t]
{\begin{center}
   \includegraphics[width=0.5\textwidth]{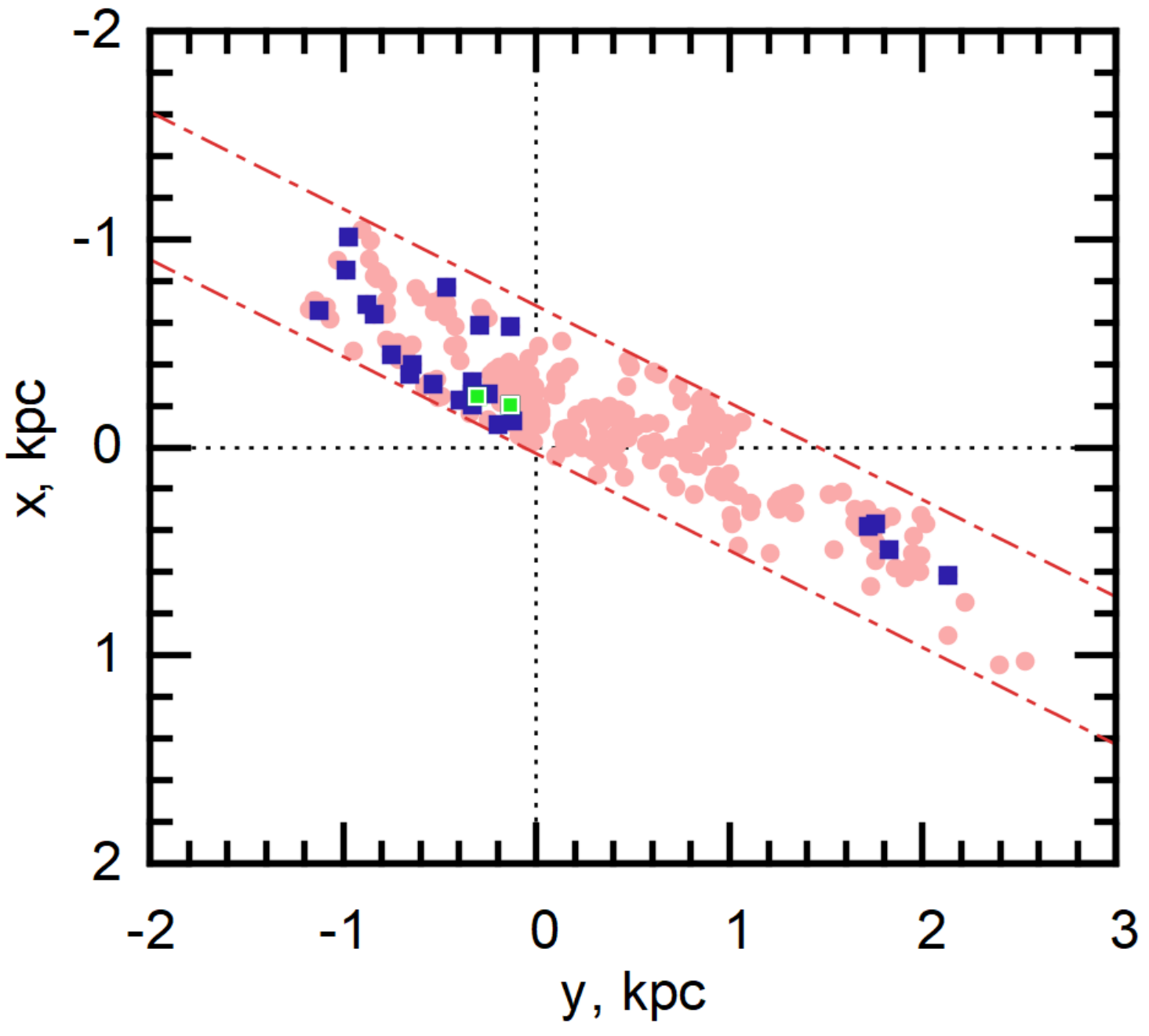}
 \caption{\small
Open star clusters younger than 70 million years in the Radcliffe Wave selection zone in the heliocentric $x,y,z$ coordinate system, see text for details.
  } \label{f-00}
\end{center}}
\end{figure}

Figure ~\ref{f-00} shows open star clusters younger than 70 Myr, selected for analysis in the Radcliffe Wave selection zone. The tilt of the selection zone to the y-axis is $-25^\circ$. Open star clusters with an age reserve were selected. For example, according to Bobylev et al. (2025b), the age of the Radcliffe Wave is approximately 30 Myr.

The proper motion data for globular clusters are taken from the new catalog by Vasiliev and Baumgardt (2021), compiled based on observations from Gaia EDR3 (Gaia Collab 2021). The average distances to globular clusters are taken from Baumgardt and Vasiliev (2021). The final catalog of globular clusters by Bajkova and Bobylev (2022), which we have at our disposal, contains 152 objects.

 \section{METHOD}
The method for solving problems of searching for open-clusters likely formed as a result of GC collisions with the Galactic disk is based on integrating the orbits of the GCs and open-clusters in the galactic gravitational potential. Integrating the orbits of the GC allows us to determine the coordinates and time of the GC orbital intersection with the Galactic disk.

In this paper, we use various rectangular galactic coordinate systems. One of them is the galactocentric $X,Y,Z$ coordinate system, where the $X$ axis points from the galactic center toward the Sun, the $Y$ axis points in the direction of galactic rotation, and the $Z$ axis points toward the north pole of the galaxy. The distance from the Sun to the galactic center is assumed to be $R_0=8.1$~kpc, which was derived as a weighted average from a large number of individual estimates in Bobylev and Bajkova (2021).

Another coordinate system is the heliocentric $x,y,z$ system, where the $x$-axis points away from the Sun toward the center of the Galaxy, the $y$-axis points in the direction of the Galaxy's rotation, and the $y$-axis points toward the North Pole of the Galaxy. We see that these two coordinate systems differ only in the direction of the $X$- and $x$-axes.

To construct the galactic orbits of the GC and OSC, we use the gravitational potential of the Galaxy $\Phi$, the axisymmetric part of which is represented as the sum of three components~--- the central spherical bulge $\Phi_b$, the disk $\Phi_d$ and the massive spherical halo of dark matter $\Phi_h$:
 \begin{equation}
 \begin{array}{lll}
  \Phi=\Phi_b+\Phi_d+\Phi_h.
 \label{pot}
 \end{array}
 \end{equation}
The bulge potentials $\Phi_b$ and disk potentials $\Phi_d$ are represented in the form proposed by Miyamoto and Nagai~(1975), and the halo component is represented according to Navarro et al.~(1997). Specific values of the parameters of the potential model used can be found in the papers of Bajkova and Bobylev (2016; 2017), where it is designated as Model~III.

The non-axisymmetric part of the potential takes into account the bar and the helical structure. The triaxial ellipsoid model was chosen as the bar potential according to the work of Palou\v{s} et al. (1993):
\begin{equation}
  \Phi_{bar} = -\frac{M_{bar}}{(q_b^2+X^2+[Ya_b/b_b]^2+[Za_b/c_b]^2)^{1/2}},
\label{bar}
\end{equation}
where $X=R\cos\vartheta, Y=R\sin\vartheta$, $a_b, b_b, c_b$~ are the three semi-axes of the bar, $q_b$~ is the length of the bar;
$\vartheta=\theta-\Omega_{bar}t-\theta_{bar}$, $tg(\theta)=Y/X$, $\Omega_{bar}$~ is the circular velocity of the bar, $t$~ is the integration time
and $\theta_{bar}$~ is the orientation angle of the bar relative to the galactic axes $X,Y$, measured from the line
connecting the Sun and the center of the Galaxy (the $X$ axis) to the major axis of the bar in the direction of the Galaxy's rotation. The following bar parameter values were adopted: $q_b=5$~kpc, $\Omega_{bar}=40$~km/s/kpc and $\theta_{bar}=25^\circ$ (Palou\v{s} et al. 1993).

In case of taking into account the spiral density wave, a term is added to the right-hand side of the expression~(\ref{pot})~(Fernandez et al. 2008):
 \begin{equation}
 \Phi_{sp} (R,\theta,t)= A\cos[m(\Omega_p t-\theta)+\chi(R)],
 \label{Potent-spir}
\end{equation}
where
 $$
 A= \frac{(R_0\Omega_0)^2 f_{r0} \tan i}{m},
 $$$$
 \chi(R)=- \frac{m}{\tan i} \ln\biggl(\frac{R}{R_0}\biggr)+\chi_\odot.
 $$
Here, $A$~ is the amplitude of the spiral wave potential,
$f_{r0}$~ is the ratio between the radial component of the perturbation from the spiral arms
and the general gravitational pull of the Galaxy,
$\Omega_p$ is the angular velocity of the rigid body rotation of the wave,
$m$~ is the number of spiral arms,
$i$~ is the arm twist angle; for a spiral pattern, $i<0$,
$\chi$~ is the phase of the radial wave; then, the phase $\chi=0^\circ$
corresponds to the center of the arm,
$\chi_\odot$~ is the radial phase of the Sun in the spiral wave.
We adopted the following values for the spiral wave parameters:
 \begin{equation}
 \begin{array}{lll}
 m=4,\\
 i=-13^\circ,\\
 f_{r0}=0.05,\\
 \chi_\odot=-120^\circ,\\
 \Omega_p=20~\hbox {km/s/kpc}.
 \label{param-spiral}
 \end{array}
 \end{equation}

\begin{figure}[t]
{\begin{center}
   \includegraphics[width=0.45\textwidth]{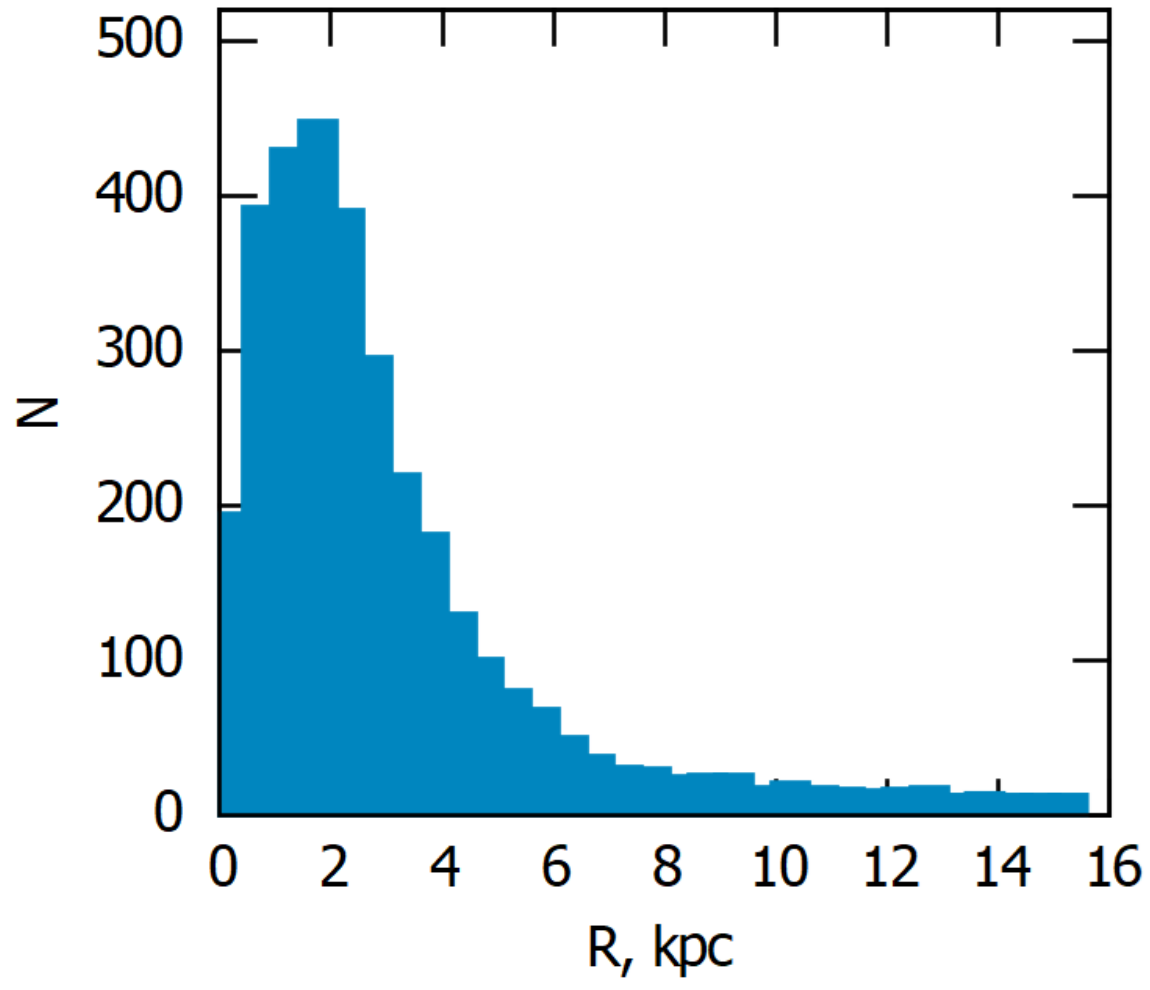}
 \caption{\small
The number of intersections, $N$, of the galactic $XY$ plane by globular clusters over the last billion years as a function of the distance to the galactic rotation axis, $R$.  }
\label{f-hist}
\end{center}}
\end{figure}
\begin{figure}[t]
{\begin{center}
   \includegraphics[width=0.99\textwidth]{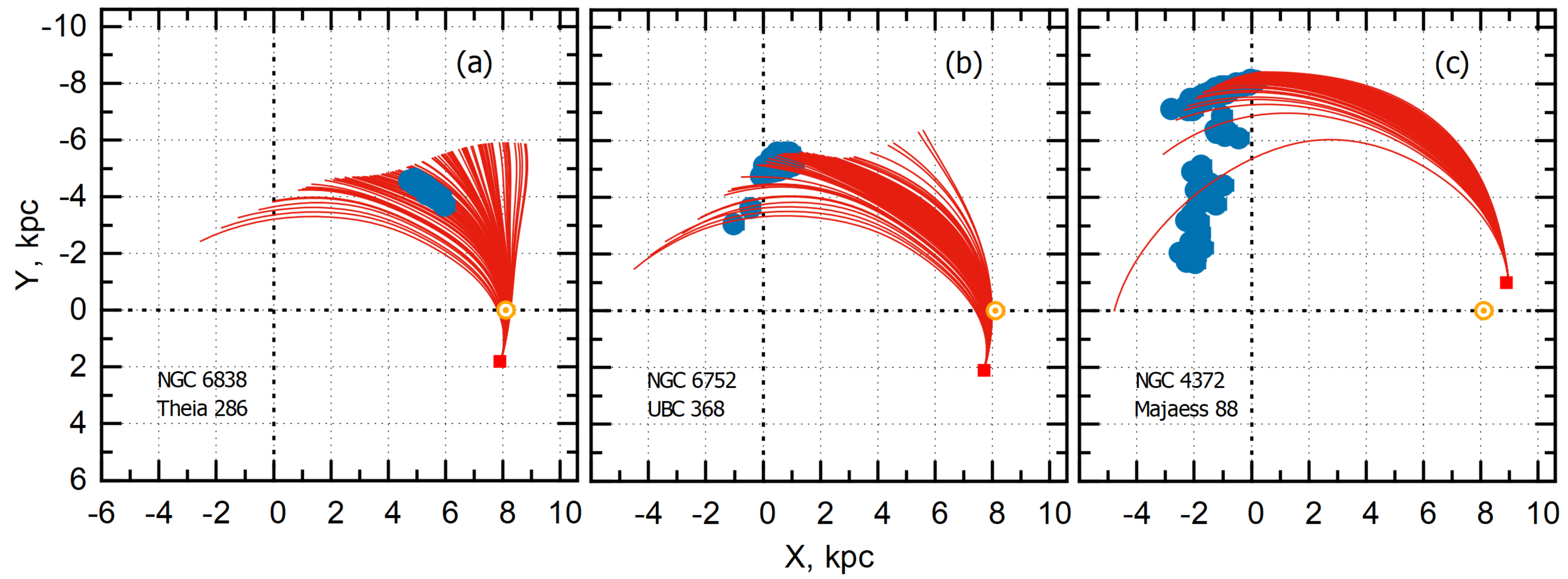}
 \caption{\small
Confidence regions of the intersection points of the Galactic $XY$ plane of globular clusters (dark blue circles) and ensembles of orbits (red lines) of open star clusters for three typical cases, the Galactic center lies at the origin of the coordinate system, the position of the Sun is marked by an orange circle. }
\label{f3}
\end{center}}
\end{figure}

The galactic orbits of GCs and OSCs are constructed in a coordinate system related to the local standard of rest. Therefore, the peculiar velocity of the Sun with values of $(U,V,W)_\odot=(11.0,12.0,7.2)$~km/s from the work of Sch\"onrich et al. (2010) is excluded from the observed initial velocities. The elevation of the Sun above the galactic plane $h_\odot=16$~pc (Bobylev and Bajkova, 2016) is also taken into account.

For each studied object (GC and OSC), statistical Monte Carlo modeling was performed. Random errors, normally distributed with zero mean and known standard deviation, were added to the object's coordinates ($X,Y,Z$), spatial velocities ($U,V,W$), and gravitational potential model parameters. The velocities, coordinates, and their errors were determined by applying Monte Carlo statistical modeling to the measured heliocentric distances, proper motions, and radial velocities of the GC and OSC.

 \section{RESULTS AND DISCUSSION}
To estimate the frequency of GC impacts on the galactic plane for each cluster, we record the moments at which $Z=0$~kpc. Most GCs have crossed the galactic disk multiple times. Overall, over the past billion years, the number of galactic plane intersections by globular clusters has been 3025. A histogram of intersections is shown in Fig. ~\ref{f-hist}. As can be seen from this figure, most of these events occur in the inner region of the Galaxy.

Fig.~\ref{f-hist} has significantly better detail compared to a similar figure from the paper
Bobylev and Bajkova (2019), which was constructed using data on 132 globular clusters and had a distribution maximum in the region of $R<1$~kpc.

	\begin{table}[t]
		\caption{The difference in times $\Delta t$ of the approach of the OSC (at the time of birth) with the place of intersection of the globular cluster NGC~4372 with the galactic plane 55.5 million years ago}
		\label{t-NGC 4372}
		\begin{center}
			\begin{tabular}{lrrrrrr}
				\hline
	           OSC name &	 Age           &  $\Delta t$    &  \\	
 	                   &	 Myr    &   Myr     &   \\\hline
  Majaess 88     &	35.2   &  20.3  &\\
    CWNU 45      &	 33.1  &   22.4 &\\
   CWNU 1024  &	 32.4  &   23.1 &\\
   CWNU 1034  &	 27.5  &   28.0 &\\
   HSC 1765      &	 38.8  &   16.7 &\\
   HSC 1808      &	 30.9  &   24.6 &\\
   HSC 1902      &	31.0   &   24.5 &\\
   HSC 1912      &	37.5   &   18.0 &\\
   HSC 1925      &	39.3  &    16.2 &\\
   HSC 1936      &	27.2   &   28.3 &\\
   LP 2383         &	37.9   &   17.6 &\\
   OC 0343        &	30.8   &   24.7 &\\
   OCSN 76       &	37.0   &   18.5  &\\
   Ruprecht 26   &	29.8   &   25.7  &\\
   Theia 35         &	 30.1  &   25.4 &\\
   Theia 399      &	26.9   &   28.6 &\\
   UPK 470        &	30.2   &   25.3 &\\
                           &	   &    &\\
        average   &	32.7   & 22.8   &\\
    \hline
			\end{tabular}
		\end{center}
	\end{table}

 \subsection{SS-RZS pairs not associated with the emergence of the Radcliffe Wave}
Fig.~\ref{f3} shows the confidence regions of the intersection points of the $XY$ galactic plane by a globular cluster and the ensembles of OSC orbits for three typical cases. Thus, Fig.~\ref{f3}a shows the pair NGC~6838 (GC)--Theia~286 (OSC); Fig.~\ref{f3}b shows the pair NGC~6752 (GC)--UBC~368 (OSC); and
Fig.~\ref{f3}c shows the pair NGC~4372 (GC)--Majaess~88 (OSC).

Let us consider in more detail the pair NGC~6838--Theia~286, shown in Fig.~\ref{f3}a. The globular cluster NGC~6838 last crossed the plane of symmetry of the Galaxy 37.5 million years ago. In this case, the angle between the direction of the GC motion and the $XY$ plane was about 10$^\circ$. The value of this angle is significant. Indeed, as shown in the works of Comer\'on and Torra~ (1992, 1994) based on numerical simulations of the flight of a massive ($3.3\times10^6M_\odot$) high-velocity cloud through the galactic disk, star formation will be most effective precisely during an oblique impactor incidence.

The age of the open star cluster Theia~286 is 7.7 million years. Therefore, the time interval $\Delta t$ between the two moments of these clusters visiting the same place was approximately 30 million years.

There is another pair for this GC with similar circumstances, namely NGC~6838--UBC~375. The age of the UBC~375 OSC is 11.0 million years, so here the time interval between the two moments of visiting the same place by the two clusters is 26.5 million years.

Consider the pair NGC~6752--UBC~368, shown in Fig.~\ref{f3}b. The globular cluster NGC~6752 last crossed the plane of symmetry of the Galaxy 41.5 million years ago. In this case, the angle between the direction of the GC motion and the $XY$ plane was about 33$^\circ$. The age of the open star cluster UBC~368 is 20.9 million years. The time interval between the two moments of visiting the same place by these clusters is approximately 20 million years.

And here there is another pair with similar circumstances, namely NGC~6752--UBC~371. The age of the OSC UBC~371 is 12.2 million years, so here $\Delta t=29.3$ million years.

Note that all four of the above-mentioned OSCs -- Theia~286, UBC~375, UBC~368, and UBC~371 -- are compactly located in Fig.~\ref{f-00} in the region of positive y-coordinate values. All of them are quite young ($t:[7.7;20.9]$~million years), younger than the Radcliffe wave, which is 30 to 40 million years old. Therefore, it is difficult to connect the intersections of the globular clusters NGC~6838 and NGC~6752 with the cause of the Radcliffe wave.

\begin{figure}[t]
{\begin{center}
   \includegraphics[width=0.99\textwidth]{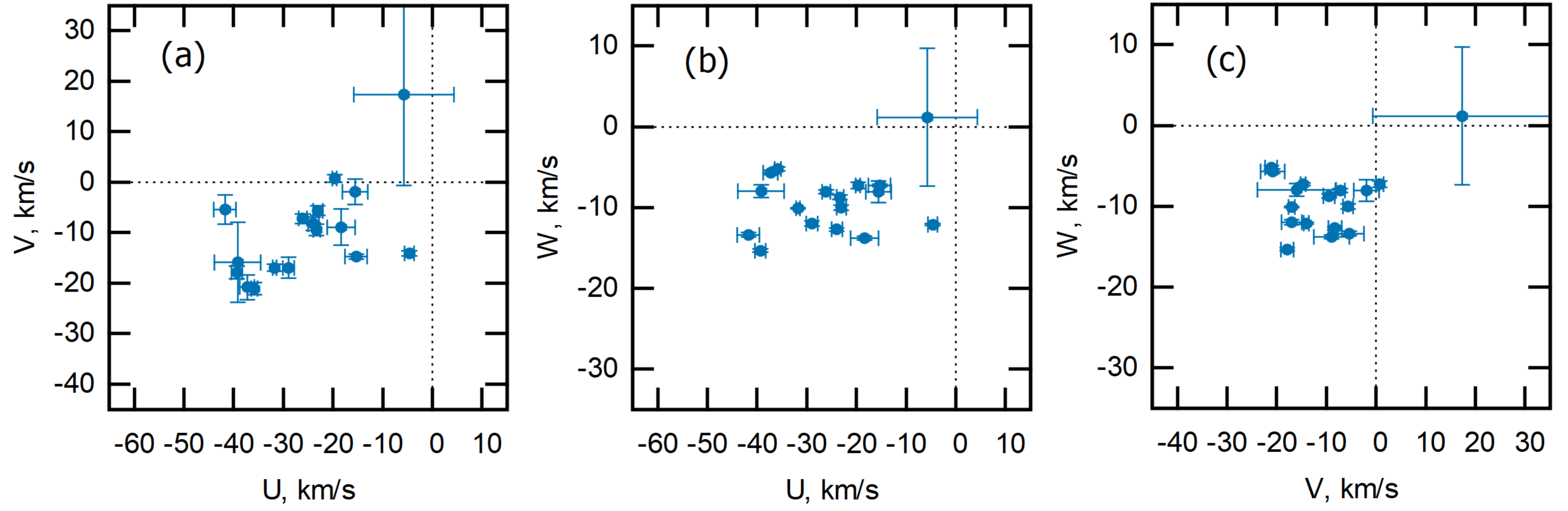}
 \caption{\small
 The velocity components $U,W,V$ and their errors for a sample of 17 OSCs listed in Table~\ref{t-NGC 4372}. }
  \label{f-UVW}
\end{center}}
\end{figure}
\begin{figure}[t]
{\begin{center}
   \includegraphics[width=0.45\textwidth]{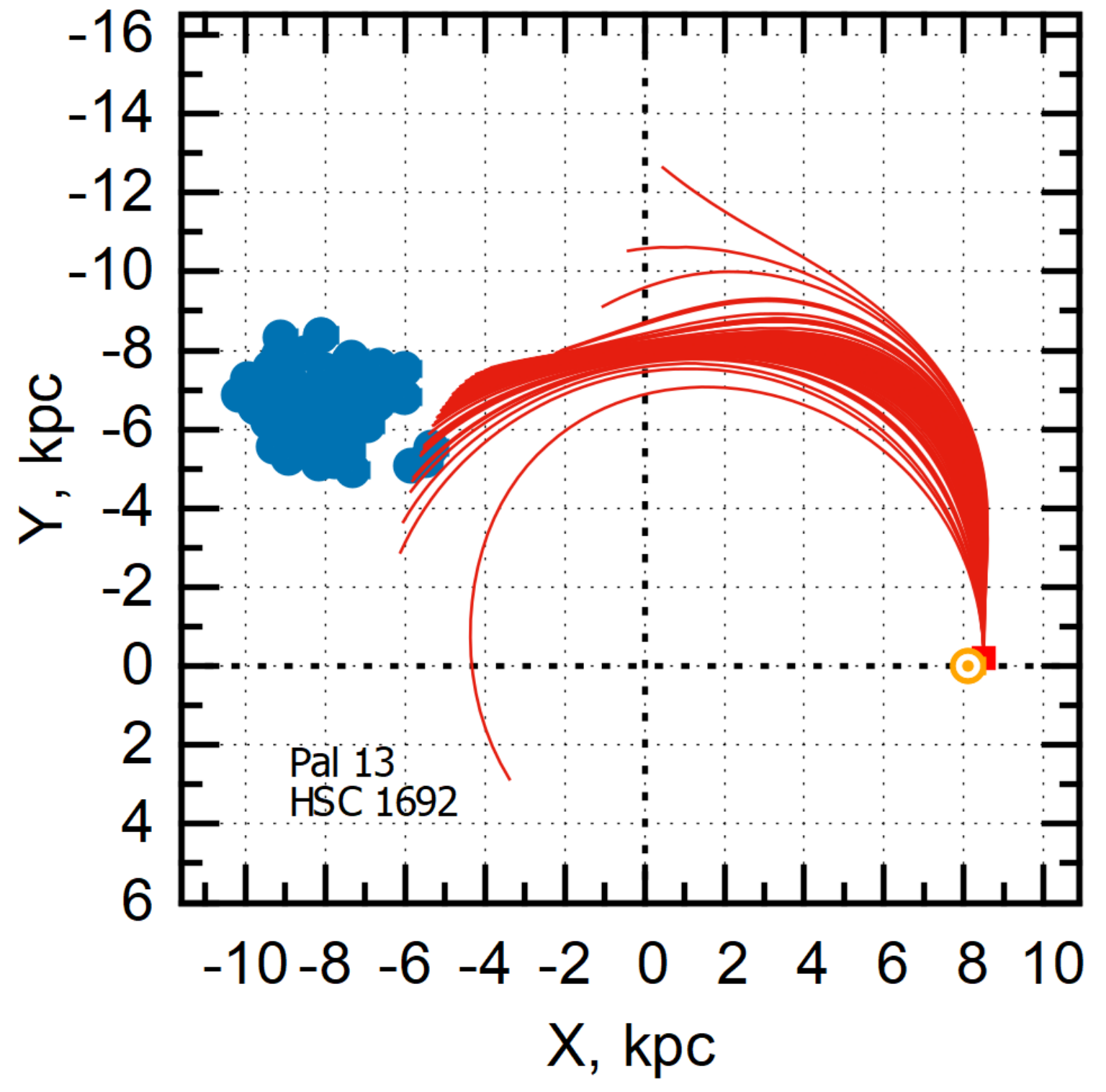}
 \caption{\small
 Confidence region of the intersection of the Galactic $XY$ plane by the globular cluster Pal 13 (dark blue circles) and the ensemble of orbits (red lines) of the OSC HSC 1692, the Galactic center lies at the origin of the coordinate system, the position of the Sun is marked by an orange circle.  }
  \label{f-Pal-13}
\end{center}}
\end{figure}

\subsection{Pairs likely associated with the occurrence of the Radcliffe wave}
There are 17 open clusters located in Fig.~\ref{f-00} in the region of negative $y$ coordinates, which are associated with the case illustrated in Fig.~\ref{f3}c. For example, the pair NGC~4372--Majaess~88, shown in Fig.~\ref{f3}c. The globular cluster NGC~4372 last crossed the plane of symmetry of the Galaxy 55.5 million years ago. In this case, the angle between the direction of GC motion and the $XY$ plane was about 33$^\circ$. The age of the open star cluster Majaess~88 is 35.2 million years, then $\Delta t= 20.3$ million years. The times of approaches of various open clusters to the intersection of the galactic plane by the globular cluster NGC~4372 are given in the table~\ref{t-NGC 4372}, where the age of the open cluster and the parameter $\Delta t$ are indicated.

The spatial velocity components $U,V,W$ of this sample of 17 OSCs are given in Fig.~\ref{f-UVW}. As can be seen from the figure, the point cloud is distributed in each of the three panels with a small dispersion. The only exception is the cluster HSC~1925, whose velocities were measured with large errors $(e_U, e_V, e_W)=(10.0, 18.0, 8.6)$~km/s.

Overall, we can conclude that the grouping of 17 open clusters is compact in coordinate space (see Fig. ~\ref{f-00}), velocity space (see Fig. ~\ref{f-UVW}), and age space (see Table ~\ref{t-NGC 4372}). This grouping of open clusters is a good candidate for supporting the hypothesis of the origin of the Radcliffe wave after the passage of an impactor through the plane of symmetry of the Galaxy. The globular cluster NGC~4372 serves as such an impactor here.

\subsection{Buy product of search}
Figure~\ref{f-Pal-13} shows the confidence region of the intersection of the Galactic plane $XY$ by the globular cluster Pal~13 85 million years ago and the ensemble of orbits of the open cluster HSC~1692. The picture is very similar for the pair Pal~13--HSC~1827, so we do not provide a separate figure for it. The positions of the two open clusters HSC~1692 and HSC~1827 in Figure~\ref{f-00} are marked with green squares. The globular cluster Pal~13 last crossed the plane of symmetry of the Galaxy 85 million years ago at an angle of $48^\circ$ to the galactic plane. The age of the open star cluster HSC~1692 is 58.0 million years, then $\Delta t=27$ million years for the pair Pal 13--HSC~1692. The age of the RZS HSC~1827 is 68.0 million years, then here $\Delta t=17$ million years.

 \section{CONCLUSION}
Based on the catalog of 152 globular clusters (GCs) from Bajkova and Bobylev (2022), orbits of the GCs were constructed to determine their intersections with the galactic plane of symmetry. Relatively young open star clusters (OSCs) were selected from the Radcliffe wave selection zone. The Hunt and Reffert (2024) catalog served as the source of data on these OSCs.

For the first time, a cluster of 17 OSCs has been discovered that is compact in coordinate, velocity, and age space. All of these open-cluster clusters, with an average age of 32.7 million years, are located in the selection zone characteristic of the Radcliffe wave. We believe this cluster grouping is a good candidate for the hypothesis that the Radcliffe wave originates from an impactor passing through the galactic plane of symmetry, with the globular cluster NGC~4372 serving as the impactor.

Indeed, the last time the globular cluster NGC~4372 crossed the Galaxy's symmetry plane was 55.5 million years ago, and 22.2 million years later, a burst of star formation occurred at this location, forming a fairly large group of open-clusters. This may have triggered the formation of the entire Radcliffe wave. These time intervals are consistent with known estimates of star formation processes. Moreover, the globular cluster NGC~4372 passed at an acute angle, approximately 33$^\circ$, to the $XY$ plane, which could have triggered the vertical oscillations in the objects' coordinates, which is the main feature of the Radcliffe wave.

As a by-product, two relatively aged OSCs, HSC~1692 (58.0 million years) and HSC~1827 (68.0 million years), were found, the formation of which could have occurred after the intersection of the plane of symmetry of the Galaxy by the globular cluster Pal~13 85 million years ago.

\bigskip{\bf BIBLIOGRAPHY}\medskip{\small
\begin{enumerate}
\item
J. Alves, C. Zucker, A.A. Goodman, et al., Nature {\bf 578}, 237 (2020).

\item
A.T. Bajkova and V.V. Bobylev, Astron. Lett. {\bf 42}, 567 (2016).

\item
A.T. Bajkova and V.V. Bobylev, Open Astronomy {\bf 26}, 72 (2017).

\item
A. T. Bajkova and V. V. Bobylev, Publ. Pulkovo Observ. 227, 1 (2022).

\item
H. Baumgardt and E. Vasiliev. Mon. Not. R. Astron. Soc. 505, 5957 (2021).

\item
V.V. Bobylev and A.T. Bajkova, Astron. Lett. {\bf 42}, 1 (2016). 

\item
V.V. Bobylev and A.T. Bajkova, Astron. Rep. {\bf 62}, 557 (2018).

\item
V.V. Bobylev and A.T. Bajkova, Astrophys. Bull. {\bf 74}, 29 (2019).

\item
V.V. Bobylev and A.T. Bajkova, Astron. Rep. {\bf 65}, 498 (2021). 

\item
V.V. Bobylev, A.T. Bajkova, and Yu.N. Mishurov, Astron, Lett, {\bf 48}, 434 (2022).

\item
V.V. Bobylev, N.R. Ikhsanov, and A.T. Bajkova, Astron. Rep. {\bf 69}, 786 (2025).

\item
V.V. Bobylev, N.R. Ikhsanov, A.T. Bajkova, Astrophys. Bull. {\bf 80}, 181 (2025b).  

\item
P. Brosche, H.-J. Tucolke, A.R. Klemola, S. Ninkovi\v{c}, M. Geffert,  and P. Doerenkamp, Astrophys. J. {\bf 102}, 2022 (1991).

\item
F. Comer\'on and J. Torra, Astron. Astrophys. {\bf 261}, 94 (1992).

\item
F. Comer\'on and J. Torra, Astron. Astrophys. {\bf 281}, 35 (1994).

\item
D. Fernandez, F. Figueras, and J. Torra, Astron. Astrophys. {\bf 480}, 735 (2008).

\item
R. Fleck, Nature {\bf 583}, 24 (2020).

\item
Gaia Collab (A. Vallenari, et al.),  Astron. Astrophys. {\bf 674}, 1 (2023). 

\item
Gaia Collab (A.G.A. Brown, et al.), Astron. Astrophys. {\bf 649}, 1 (2021). 

\item
E.L. Hunt and S. Reffert, Astron. Astrophys {\bf 696}, A42 (2024).

\item
R. Konietzka, A.A. Goodman, C. Zucker, et al.,  Nature {\bf 628}, 62 (2024).

\item
V.V. Levy, Astron. Astrophys. Trans. {\bf 18}, 621 (2000).

\item
A. Marchal and P.G. Martin, Astrophys. J. {\bf 942}, 70 (2023).

\item
R. de la Fuente Marcos, C. de la Fuente Marcos, and D. Reilly,  Astrophys. Space Science {\bf 349}, 379 (2014).

\item
M. Miyamoto and R. Nagai, Publ. Astron. Soc. Japan {\bf 27}, 533 (1975).

\item
J.F. Navarro, C.S. Frenk, and S.D.M. White,  Astrophys. J. {\bf 490}, 493 (1997).

\item
J. Palou\v{s}, B. Jungwiert, and J.~Kopeck\'y,  Astron. Astrophys. {\bf 274}, 189 (1993). 

\item
R.F. Rees and K.M. Cudworth,  Bulletin of the American Astronomical Society {\bf 35}, 1219 (2003).

\item
R. Sch\"onrich, J. Binney, and W. Dehnen,     Mon. Not. R. Astron. Soc. {\bf 403}, 1829 (2010).

\item
L. Thulasidharan, E. D'Onghia, E. Poggio, et al.,   Astron. Astrophys. {\bf 660}, 12 (2022).

\item
D. Vande Putte and M. Cropper, Mon. Not. R. Astron. Soc. {\bf 392}, 113 (2009).

\item
E. Vasiliev and H. Baumgardt, Mon. Not. R. Astron. Soc. 505, 5978 (2021).

\item
J.F. Wallin, J.L. Higdon, and L. Staveley-Smith, Astrophys. J. {\bf 459}, 555 (1996).

\end{enumerate} }

\end{document}